\documentclass[aps,prd,amsfonts,amsmath,nofootinbib,tightenlines,preprint,showpacs,longbibliography,superscriptaddress]{revtex4-2}
\pdfoutput=1

\usepackage{graphicx}
    \graphicspath{ {./figures/} }
\usepackage{amsmath}
\usepackage{amssymb}
\usepackage{mathtools}
\usepackage{array}
\usepackage[caption=false]{subfig}

\def\be{\begin{equation}}
\def\ee{\end{equation}}

\newcommand{\Gref}{\Gamma_{\!*}}
\newcommand{\GGmu}{\Gamma_{\!*} G\mu}
\newcommand{\GGmux}{\Gamma G\mu}

\begin{document}

\title{More accurate gravitational wave backgrounds from cosmic strings}

\author{Jeremy M. Wachter}
\email{wachterj@wit.edu}
\affiliation{School of Sciences \& Humanities, Wentworth Institute of Technology, Boston, MA 02155, USA}

\author{Ken D. Olum}
\email{kdo@cosmos.phy.tufts.edu}
\affiliation{Institute of Cosmology, Department of Physics and Astronomy, Tufts University, Medford, MA 02155, USA}

\author{Jose J. Blanco-Pillado}
\email{josejuan.blanco@ehu.es}
\affiliation{IKERBASQUE, Basque Foundation for Science, 48011, Bilbao, Spain}
\affiliation{Department of Physics, University of the Basque Country UPV/EHU, 48080, Bilbao, Spain}
\affiliation{EHU Quantum Center, University of the Basque Country UPV/EHU, Bilbao, Spain}

\begin{abstract}

We derive a general procedure for calculating the gravitational wave background (GWB) from cosmic string loops whose typical shape evolves over time, as in gravitational backreaction. Using the results of a large-scale study of numerical gravitational backreaction on Nambu-Goto cosmic string loops, we construct GWBs of backreacted cosmic strings for a range of tensions and frequencies of cosmological interest, and compare them to current and upcoming gravitational wave detectors. The GWBs are lower than prior predictions by anywhere from a few percent to around 30\%, depending on the frequency and tension in question.

\end{abstract}

\maketitle
\newpage

\section{Introduction}\label{sec:intro}

The dawn of the era of gravitational wave astronomy for both transient and background signals~\cite{PhysRevLett.116.061102,NANOGrav:2023gor,Antoniadis:2023rey,Reardon:2023gzh,Xu:2023wog} has increased the need for precise templates for gravitational wave sources. Currently-operating gravitational wave detectors require them for performing searches as well as subtracting out confusion noise to more precisely measure other signals. Future detectors need them for planning their data collection and analysis pipelines, as well as predicting what sort of physics they'll be able to measure or constrain. The more precise these templates are, the better science can be done; while most potential gravitational-wave sources have well-developed predictions for their signals, not all of the details are filled in. In the case of cosmic strings, the effect of gravitational self-interactions, or \emph{backreaction}, on the loop gravitational wave signals has only been accounted for in toy models~\cite{Blanco-Pillado:2017oxo,Blanco-Pillado:2021ygr}. In this work, we will present the gravitational-wave background (GWB) of a cosmic string network accounting for backreaction with accurate numerical methods. These methods describe the time-dependent behavior of the power radiated into gravitational waves by mode, an improvement over toy models which assume a smoothed loop's power is fixed over the course of its lifetime.

Gravitational waves from cosmic strings are an important category of signals to search for. Strings are common predictions of theories beyond the standard model of particle physics~\cite{Jeannerot:2003qv,Lazarides:2019xai}, forming either in a field theory as topological defects generated during a spontaneous symmetry breaking or as fundamental strings in string theory~\cite{Vilenkin:2000jqa,Sarangi:2002yt,Dvali:2003zj}. They produce gravitational waves (GWs) and, because a network of long strings and loops forms in the early universe and then quickly reaches a scaling solution~\cite{Vilenkin:1981kz,Kibble:1984hp,Vanchurin:2005pa,Blanco-Pillado:2013qja},\footnote{Changes to the Universe's equation of state can temporarily disrupt this scaling solution; see~\cite{Sousa:2013aaa,Gouttenoire:2019kij} for recent discussions.} they produce a GWB which spans a large range of frequencies. Detecting a field theory string or string network is informative for particle physics because it sets the scale of a phase transition, and detecting a fundamental string would be a confirmation of string theory (although additional measurements, such as a measurement of reconnection probability below $1$, would be necessary to confirm that the object is string-theoretical).  Detecting a string GWB can also be informative for cosmology and astrophysics, as the GWB contains the imprint of the large-scale processes and evolution of the universe.

We will work here with cosmic strings interacting only with gravity in the Nambu-Goto (infinitely-thin) limit and assuming a reconnection probability of $1$. Then the strings are characterized by a single parameter, the tension $\mu$. For gravitational physics, the dimensionless\footnote{We work in units where the speed of light is 1} parameter $G\mu$ is more typically used. Current non-observation of string GWBs in pulsar timing arrays (PTAs) sets the strongest upper bound on the string tension of $G\mu\lesssim 10^{-10}$~\cite{NANOGrav:2023hvm}.

The paper is laid out as follows. In Sec.~\ref{sec:setup}, we review how our loop population synthesis and gravitational backreaction code work; we also introduce important terms and definitions for this work. In Sec.~\ref{sec:gwb}, we develop a new formalism for finding the energy density in cosmic string loops assuming that the power radiated into gravitational waves changes over the loop's lifetime. In Sec.~\ref{sec:results}, we show the results of applying this new formalism to our numerically-evolved loops: gravitational-wave backgrounds for string loops accounting for gravitational backreaction. We comment on detectability by current and future gravitational-wave detectors, compare these new results to prior results, and comment on possible refinements to produce even more accurate gravitational wave backgrounds from cosmic strings. We conclude in Sec.~\ref{sec:conc}.

\section{Setup}\label{sec:setup}

\subsection{Numerical gravitational backreaction}\label{ssec:backreaction}

Our backreaction code builds on the code base of Refs.~\cite{PhysRevD.95.023519,Blanco-Pillado:2019nto}; for fuller details, see Ref.~\cite{results-paper}.  Techniques of this kind were first used by Allen and Casper~\cite{Allen:1994iq}.  The essential ideas we summarize here. 

In the Nambu-Goto (NG) limit, a string is a one-dimensional object, and so sweeps out a worldsheet in spacetime. We put down null parameters $u$ and $v$ on this worldsheet and describe the string's position using the worldsheet functions $A(v)$ and $B(u)$:
\be
    X^\gamma(u,v) = \frac12\left[A^\gamma(v)+B^\gamma(u)\right]\,,
\ee
which represent the general, unperturbed solution of the NG equations of motion in the conformal gauge, namely, $X^\gamma_{,uv} = 0$. In the presence of a gravitational field, the string's position is modified according to 
\be\label{eqn:Xuv}
    X^\gamma_{,uv} = \frac14\Gamma^\gamma_{\alpha\beta}A'^\alpha B'^\beta\,,
\ee
where $\Gamma^\gamma_{\alpha\beta}$ is the Christoffel symbol. The spatial components of this acceleration term tell us how the shape of the string changes, while the time component tells us how energy is lost into gravitational waves.

Looking only at non-self-intersecting string loops, we can separate oscillatory effects from secular effects by allowing the total worldsheet correction due to gravitational interactions to accumulate over many oscillations, $N_\text{osc}$, of the loop (while still keeping $N_\text{osc}G\mu$ small) without changing its shape. Then, we apply the changes all at once, re-calculate the gravitational interactions for the new shape, and repeat the process until our loop has evaporated to some target fraction of its initial length. Because $G\mu\ll 1$, we may work in linearized gravity.

In order to represent our loops numerically, we use a piecewise-linear model. Here $A(v)$ and $B(u)$ consist of sequences of straight segments, so $A'(v)$ and $B'(u)$ are sequences of constant null four-vectors; when applying the accumulated effects of backreaction as mentioned above, we rotate and dilate these segments to maintain the piecewise-linear model, although a real piecewise-linear string would have its straight segments bent by backreaction (see, e.g., Fig. 10 of Ref.~\cite{PhysRevD.95.023519}). We additionally use our freedom to rescale the $u,v$ null coordinates to ensure that these null four-vectors always have unit time component. This significantly simplifies the calculation of the Christoffel symbol and backreaction four-acceleration term (see \cite{PhysRevD.95.023519} for derivations of the relevant equations and~\cite{Vilenkin:2000jqa} for a detailed discussion on string worldsheet parametrizations). In order to achieve faithful representations of structure evolution on loops, our code has been designed to handle backreaction computations between segments of very different sizes. In addition, we introduce finer discretization of any segment which forms a large spatial angle with its successor segment.

Lastly, when discussing gravitational effects of loops, an essential measure is the gravitational wave power of a loop, $\Gamma$. This is a geometric quantity (depending only on the shape of the loop) which gives the rate of loop length lost into gravitational radiation as $dL/dt=-\Gamma G\mu$. We compute $\Gamma$ by the procedure described in Sec. III.C of~\cite{results-paper}.

\subsection{The loop corpus}\label{ssec:corpus}

Reference~\cite{results-paper} reports on the details of the loop populations we evolved under backreaction. For this work, the most important loops are what that reference termed the \emph{70\% evaporated no-majors sub-population}. These are the 71 loops which were evaporated until 70\% of their starting length was lost to backreaction (i.e., gravitational wave emission), and which also experienced at most minor self-intersections over the course of their evolution.\footnote{All of the loops we took from our network simulations were non-self-intersecting; however, as reported in Ref.~\cite{results-paper}, backreaction can lead to (typically minor) self-intersections of loops.} We chose to exclude 70\%-evaporated loops which experienced larger length losses to self-interactions in order to focus primarily on spectral shape changes due to backreaction processes. When we discuss the power spectrum of loops, $P_n$, and how it evolves due to backreaction, it is these loops' power spectra we will use as our basis. The shape and evolution of $P_n$ plays a central role in subsequent results, and so too does its modeling (and approximations made therein); see Ref.~\cite{results-paper} for more details, and Sec.~\ref{ssec:discuss} for a discussion of how to improve this aspect of the model. The binned $nP_n$ data are available at Ref.~\cite{wachter_2024_14037539}.

Because we use a piecewise-linear model, one way to characterize our loops is by their \emph{segmentation}, or the number of segments in $A(v)$ plus the number of segments in $B(u)$. The segmentation value sets an upper bound on the number of kinks---discontinuous changes in the loop's tangent vectors---we could say this loop contains. The ``true'' number of kinks could always be lower---if, for example, a series of tiny changes in angle represents a curved section of string, and not a sequence of kinks---but we have no method of definitively separating ``true'' kinks from ``discretization'' kinks.

The population of loops we studied has a typical total segment count on the order of several hundred. Thus, our simulations track the evolution of loops with several hundred kinks. Real loops will generally have more kinks due to a longer period of network evolution than what can be simulated; we address how to connect our results to the GWB expected from loops in nature in Sec.~\ref{ssec:discuss}.

\subsection{Measuring loop ages and lengths}\label{ssec:ages}

To calculate the gravitational wave background of string loops as measured today, we must integrate over space and time. To find the correct contribution to the total gravitational wave spectrum at each point in time, we must know how the loop shapes have been affected by backreaction. We are therefore interested in relating the length and age of a loop to the amount of backreaction that loop has undergone.

Let us begin with quantifying how long a loop has existed. There are a number of ways to measure this; for example, if we consider a loop created at some physical time $t'$, then at a time $t$, the loop has existed for time $t-t'$. While this measure is straightforward, it is not particularly useful when thinking about the scale of backreaction, which depends critically on the length of the loop (and thus the number of oscillations it has completed) and the strength of the string's coupling to gravity. Longer loops or smaller couplings should both lead to a slower build-up of backreaction effects. So inspired, we introduce the \emph{normalized loop age},
\be\label{eqn:zeta-def}
    \zeta = \frac{\Gref G\mu (t-t')}{L'}\,,
\ee
where $L'$ is the invariant length of the loop (its energy in units of $\mu$) at its time of creation and $\Gref$ is the $\Gamma$ of a ``canonical'' loop, which here we take to be $50$. Nothing depends on the value of $\Gref$; it is used only to set a benchmark and for comparison to other work. A loop with constant $\Gamma$ evolves as $L=L'-\Gamma G\mu(t-t')$, and so has a lifetime $t_\text{life} = L'/(\Gamma G\mu)$; thus, we can understand $\zeta$ as being normalized against the lifetime of a loop with a constant $\Gamma=\Gref$. This lets us compare loops of different initial sizes within a population to each other, as well as compare loops in a tension-independent way. While all loops clearly start with $\zeta=0$, it is less clear what $\zeta$ they will have at the end of their lives. We will return to this idea in a moment.

Another measure, discussed at length in~\cite{results-paper}, is the evaporation fraction,
\be\label{eqn:chi-def}
    \chi = 1-\frac{L}{L'}\,,
\ee
where $L$ is the current invariant length of a loop. Manifestly, $\chi$ runs from $0$ to $1$. However, the relationship of $\chi$ to time is not clear if the rate at which a loop's length changes---the rate at which it loses energy into gravitational waves---changes over time. That $\Gamma$ does change over time is a key finding of~\cite{Blanco-Pillado:2019nto,results-paper}.

Let us now connect these two measures. The dimensionless measure of loop power, $\Gamma$, is related to the actual power emitted into gravitational waves by $P_\text{gw}=G\mu^2\Gamma$. From $dE/dt=P$ and $E=\mu L$, we can write $dL/dt=-G\mu\Gamma$. Allowing $\Gamma$ to be a function of the normalized age and transforming variables $t\rightarrow\zeta$, we have
\be\label{eqn:L-fxn-zeta}
    L = L'\left(1 - \int^\zeta_0\frac{\Gamma(\zeta')}{\Gref}\,d\zeta'\right)\,,
\ee
where $\Gamma(\zeta)$ has support on $\zeta=[0,\zeta_\text{max}]$, and $\zeta_\text{max}$ is defined by
\be\label{eqn:Gref-integral}
    \Gref = \int^{\zeta_\text{max}}_0\Gamma(\zeta')\,d\zeta'\,.
\ee
A loop with $\zeta_\text{max}>1$ ``lives longer'' than the reference loop due to a lower (on average) $\Gamma$, and vice versa. In the case of constant $\Gamma$, $\zeta = (\Gref/\Gamma)\chi$ and $\zeta_\text{max}=\Gref/\Gamma$.

In addition, we can combine Eqs.~(\ref{eqn:chi-def},\ref{eqn:L-fxn-zeta}) to find
\be\label{eqn:chi-fxn-zeta}
    \chi = \int^{\zeta}_0\frac{\Gamma(\zeta')}{\Gref}\,d\zeta'\,.
\ee
Thus, if we understand how $\Gamma$ changes with the normalized loop age, we can connect the evaporation fraction and normalized loop age. It is a useful observation from the above equation that $d\chi/d\zeta = \Gamma/\Gref$.

\begin{figure}
    \centering
    \includegraphics[scale=1.00]{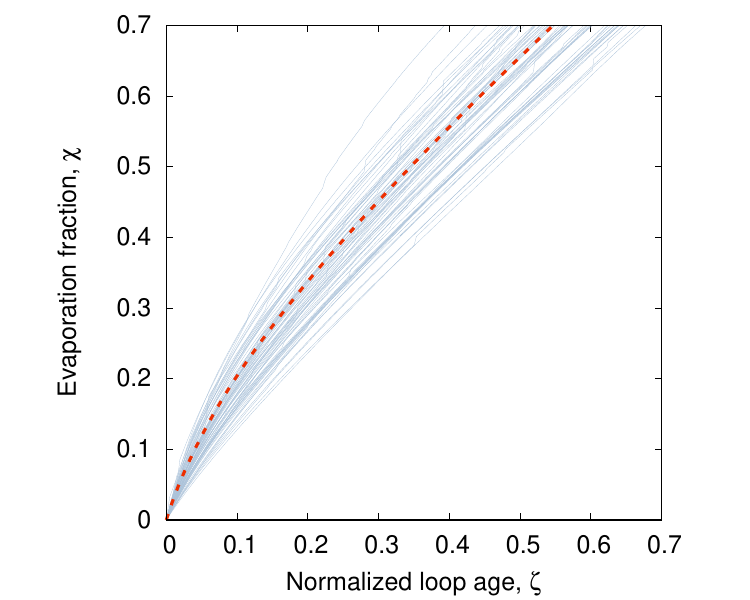}
    \caption{The dependence of evaporation fraction on normalized loop age. The individual results for 71 loops are shown as thin solid blue lines, and the average $\chi$ is shown as a dashed red line. The slope of the red line lets us find a functional form for $\Gamma(\zeta)$. Larger $\Gamma$ at earlier times means that loops evaporate more rapidly as compared to later times, where the $\chi$-$\zeta$ relationship approaches a straight line.}\label{fig:age-vs-evap}
\end{figure}

Reference~\cite{results-paper} found that loops tend to start with a larger $\Gamma$, which decreases, fairly quickly to begin with, until it approaches an asymptotic value a bit below the canonical $\Gamma\sim 50$. Loops are therefore younger when they reach a particular evaporation fraction than they would be if they had evolved with a fixed, canonical $\Gamma$. This can be seen in Fig.~\ref{fig:age-vs-evap}, where we construct a plot of $\chi$ versus $\zeta$ for all loops in our 70\%-evaporated no-majors subpopulation.

The larger initial slope indicates that loops evaporate more rapidly in their youth, eventually tending towards a lower (and constant) rate of evaporation.  We can read the instantaneous $\Gamma$ of a loop off of the slope of these lines. A linear relationship implies a constant $\Gamma$. The averaged results indicate $\Gamma\approx47$ at large $\chi$, consistent with the asymptotic behavior of the average $\Gamma$ of this population of loops found in~\cite{results-paper}.

Once we reach the end of our data, we are faced with a choice. While our results indicate that the loop population is tending towards a constant average $\Gamma$, it has not reached that value for the 70\%-evaporated no-majors sub-population. Yet we need a way to extend our results out to $\chi=1.0$.\footnote{We discuss the computational complexity of evaporating to higher $\chi$ in~\cite{results-paper}: going to larger $\chi$ quickly yields diminishing returns and reaching $\chi=1.0$ is impossible, so extrapolation is needed at some point regardless.} The simplest idea is to make a linear extrapolation at some fixed $\Gamma$. We have elected to take the $\Gamma\approx47$ we fit above and treat that as the asymptotic constant, extrapolating the average $\chi(\zeta)$ as linear in $\zeta$ for $\chi>0.7$ (i.e., we extend the red line in Fig.~\ref{fig:age-vs-evap} on a straight trajectory until we reach $\chi=1.0$). This over-estimates the rate of evaporation; physically, it corresponds to freezing the loop shape at this stage, while the loop still shrinks by emitting energy into gravitational waves.

Extrapolating with $\Gamma\approx 47$ yields $\zeta_\text{max}\approx 0.88$. Loops evolved with a more accurate model of backreaction are typically shorter-lived than canonical loops, though not by much.\footnote{$\Gamma\approx 47$ is an upper bound on the rate of radiation, since we do not find any regime in which the average loop $\Gamma$ increases. As a lower bound, we might take the conjectured minimum $\Gamma\approx39$~\cite{PhysRevD.50.3703,0264-9381-22-13-002}. This is less physical, since it implies that at around $\chi=0.7$, every loop suddenly transforms into the minimum-radiator ACO~\cite{Allen:1994bs} loop. With this choice, we would find $\zeta_\text{max}\approx 0.95$.}

\section{Calculation of the gravitational wave background}\label{sec:gwb}

\subsection{The loop density function}\label{ssec:loop-density}

Assume we have some loop production function $\mathsf f(L',t')$, such that $\mathsf f(L',t')dL'dt'$ is the number of loops with lengths $[L',L'+dL']$ created at time $[t',t'+dt']$ per unit physical volume. We wish to obtain a loop density function $\mathsf n(L,\zeta,t)$ such that $\mathsf n(L,\zeta,t)dLd\zeta$ is the number of loops per physical volume with lengths $[L,L+dL]$ and ages $[\zeta,\zeta+d\zeta]$ existing at time $t$.

Let us first ignore length loss to gravitational radiation as well as energy loss to redshifting of loop velocities. Then, the only change to the loop number density in a \emph{comoving} volume is due to the production of loops. Thus,
\be
    \frac{d}{dt}\left[a^3(t)\mathsf n(L,t)\right] = a^3(t)\mathsf f(L,t)\,,
\ee
where $a(t)$ is the scale factor of the universe. More generally,
\be\label{eqn:n-integral}
    \mathsf{n}(L,\zeta,t) = \int^t_0dt'\int^\infty_0dL'\left(\frac{a(t')}{a(t)}\right)^3\mathsf f(L',t')\delta\left(L-L(L',t',t)\right)\delta\left(\zeta - \zeta(L',t',t)\right)\,,
\ee
where Dirac delta functions enter under the assumption that all loops age in the same way (or: all loops have the same $\Gamma(\zeta)$). Once we know the length and time of formation as well as the current time, the current length and age are determined.

Let us rewrite the Dirac delta functions in a more amenable form for integration. We use the definition of $\zeta$ from Eq.~(\ref{eqn:zeta-def}) for $\zeta(L',t',t)$, and obtain $L(L',t',t) = L'(1-\chi)$ from Eq.~(\ref{eqn:chi-def}) (where we note that $\chi$ is a function of $\zeta$ only, as given by Eq.~(\ref{eqn:chi-fxn-zeta})). Then,
\be\label{eqn:double-delta-transform}
    \delta\left(L-L(L',t',t)\right)\delta\left(\zeta - \zeta(L',t',t)\right) = \delta(L'-\bar{L}')\delta(t'-\bar{t}')\left|\mathcal{J}(L',t')\right|^{-1}\,,
\ee
where $\bar{L}',\bar{t}'$ are the values such that $L-L(\bar{L}',\bar{t}',t)=0$, $\zeta-\zeta(\bar{L}',\bar{t}',t)=0$, and $\mathcal{J}(L',t')$ is the Jacobian for transforming from $(L',t')$ coordinates to $(L,\zeta)$ coordinates. As it is easier to write, we will preferentially work with the inverse transformation, $\mathcal{J}(L,\zeta)$, and use the property that $\mathcal{J}(L,\zeta)=\mathcal{J}(L',t')^{-1}$.

To find $\left|\mathcal{J}(L,\zeta)\right|$, we rearrange Eqs.~(\ref{eqn:zeta-def},\ref{eqn:chi-def}) to get
\be\label{eqn:L0-t0}
  L' = \frac{L}{1-\chi}\,,\qquad t' = t-\frac{L\zeta}{\Gref G\mu (1-\chi)}\,.
\ee
These functions, plus the observation that $\partial\chi/\partial\zeta=\Gamma/\Gref$, allow us to construct the Jacobian:
\be\label{eqn:jacobian}
    \left|\mathcal{J}(L,\zeta)\right| = \left|\begin{array}{cc} \frac{\partial L'}{\partial L} & \frac{\partial L'}{\partial \zeta} \\ \frac{\partial t'}{\partial L} & \frac{\partial t'}{\partial \zeta} \end{array}\right| = \left|\begin{array}{cc} \frac{1}{1-\chi} & \frac{\Gamma}{\Gref}\frac{L}{(1-\chi)^2} \\ -\frac{\zeta}{\Gref G\mu(1-\chi)} & -\frac{L}{\Gref G\mu(1-\chi)}\left(1+\frac{\Gamma}{\Gref}\frac{\zeta}{1-\chi}\right) \end{array}\right| = -\frac{L}{\Gref G\mu(1-\chi)^2}\,.
\ee
Note that this is singular as $\chi\rightarrow 1$ (and, simultaneously, $L\rightarrow 0$). This has the physical interpretation that, at any time, the loops which have just finished fully evaporating could have come from a huge range of $(L',t')$ pairs, from ``very small and recent'' to ``very large and long ago''.

Finally, we observe that $a(t)\propto t^\nu$, with $\nu=1/2,2/3$ in the radiation and matter eras, respectively. We assume for now that $\nu$ is fixed, but will address the effect of changing cosmological era further on.

Putting this all together, we integrate Eq.~(\ref{eqn:n-integral}) and obtain
\be
    \mathsf n(L,\zeta;t) = \frac{L}{\Gref G\mu(1-\chi)^2}\left(1-\frac{L\zeta}{\Gref G\mu(1-\chi)t}\right)^{3\nu}\mathsf f\left(\frac{L}{1-\chi},t-\frac{L\zeta}{\Gref G\mu(1-\chi)}\right)\,.
\ee
This is as far as we can go in deriving a form for $\mathsf n(L,\zeta;t)$ without a specific form for $\mathsf f(L',t')$. However, we can do some further analysis to determine which $L$ and $\zeta$ are most important by writing the loop production in scaling form, $\mathsf f(x')$, with $x'=L'/t'$ the scaling size at production (and $x=L/t$ the scaling size more generally). For any specific $(\zeta,t,x')$ triplet, the length of loop which is relevant is
\be\label{eqn:L-dependence}
    L = \frac{\Gref G\mu t x' (1-\chi)}{x'\zeta + \Gref G\mu}\,.
\ee
Equation~(\ref{eqn:L-dependence}) answers the following question: Given a loop with normalized age $\zeta$ and corresponding evaporation fraction $\chi$, how long must this loop be to have been created with fixed $x'$ and reached the given $\zeta$ at time $t$?  The resulting $L$ is proportional to $t$, as required for scaling.  So long as $x'\zeta\gg \Gref G\mu$, $L$ is effectively independent of $x'$ and has a simple monotonically-decreasing relationship on $\zeta$.  In this regime, the age of the loop is essentially $t$, and $L$ must be small enough to reach $\zeta$ in that time.

But if we go to a very small $\zeta$, $L$ does not increase indefinitely, because of the second term in the denominator of Eq.~(\ref{eqn:L-dependence}), which results from the presence of $t'$ in Eq.~(\ref{eqn:zeta-def}).  Instead around $x'\zeta\approx \Gref G\mu$, the increase tapers off and $L\rightarrow (1-\chi)tx'=L'$ as $\zeta\rightarrow 0$, as expected.  Such loops have formed very recently, with $t'$ not much less than $t$.

Studying loops in terms of their scaling size is useful more generally as well. Let's back up to the point where we had Eqs.~(\ref{eqn:n-integral},\ref{eqn:double-delta-transform},\ref{eqn:jacobian}). Before performing the integral, let's convert the loop production function and loop number density to scaling units:
\be
    \mathsf n(x,\zeta) = t^4\mathsf n(L,\zeta;t)\,,\qquad \mathsf f(x') = t'^5\mathsf f(L',t')\,.
\ee
Thus,
\be\label{eqn:n-integral-alt}
    \mathsf{n}(x,\zeta) = \int^t_0dt'\int^\infty_0dL'\left\{\frac{x}{\GGmu(1-\chi)^2}\left(\frac{t}{t'}\right)^{5-3\nu}\mathsf f\left(\frac{L'}{t'}\right)\delta(L'-\bar{L}')\delta(t'-\bar{t}')\right\}\,.
\ee
Carrying out the two integrals, we find
\be\label{eqn:n-x-zeta}
    \mathsf n(x,\zeta) = \frac{\GGmu x\left((1-\chi)\GGmu\right)^{3-3\nu}}{((1-\chi)\GGmu-x\zeta)^{5-3\nu}}\mathsf f\left(\frac{\GGmu x}{(1-\chi)\GGmu-x\zeta}\right)\,.
\ee
This expression is complete assuming that we are always in the same cosmological era---a fixed $\nu$. However, loops created in the radiation era but emitting in the matter era will be diluted due to the change in the rate of expansion of the universe. To account for this, we will multiply the loop number density by an additional dilution factor $\mathcal{S}(z,\zeta)$, derived in Appendix~\ref{app:relic-loops}.

The loop number density of Eq.~(\ref{eqn:n-x-zeta}) is manifestly non-separable, with non-trivial interactions between $x$ and $\zeta$. Adapting Eq.~(\ref{eqn:L-dependence}), we find that the relevant $x$ given some $(\zeta,x')$ pair is
\be\label{eqn:x-dependence}
    x = \frac{\GGmu x' (1-\chi)}{x'\zeta + \GGmu}\,.
\ee
This correctly goes to $x'$ when $\zeta=0$ and to $0$ when $\zeta=\zeta_\text{max}$ (thus $\chi=1$). In-between, we see a similar effect as in $L$ where $x$ decreases less rapidly than linearly for very small $\zeta$.\footnote{We can recover the usual, age-independent loop number density---see, e.g., Ref.~\cite{Blanco-Pillado:2013qja}---by integrating over $\zeta$. See Appendix~\ref{app:original-n} for details.}

\subsection{The density of gravitational waves}\label{ssec:gw-density}

We will describe the gravitational wave background using a dimensionless measure: the energy density of gravitational waves in unit logarithmic frequency intervals, in units of the critical density:
\be\label{eqn:gwb}
    \Omega_\text{gw}(\ln f) = \frac{8\pi G}{3H_0^2}f\rho_\text{gw}(t_0,f)\,,
\ee
where $t_0$ is the current time, $H_0$ the current Hubble parameter, and $\rho_\text{gw}$ the energy density of gravitational waves per unit frequency. This energy density is calculated by integrating the redshifted total energy deposited across cosmic time at each frequency $f=f'/(1+z)$, where $f'$ is the frequency of emission and $f$ the frequency we measure today.

At each point in time we must add up the GW power of all extant loops.  Prior work separates the power spectrum and cosmological effects into separate terms, and thus writes something like Ref.~\cite{Blanco-Pillado:2017oxo}: $\rho_\text{gw}(t_0,f) = G\mu^2\sum_{n=1}^\infty C_n(f)P_n$, where $C_n(f)$ is the result of an integral across cosmic time. However, as seen in Eq.~(\ref{eqn:n-x-zeta}), we cannot assume a power spectrum independent of the loop history. We model the GW emission from loops as arising from the power spectrum $P_n(\zeta)$ computed using the methods described in Ref.~\cite{results-paper} and available at Ref.~\cite{wachter_2024_14037539}, and write
\be
    f\rho_\text{gw}(t_0,f) = G\mu^2\sum_{n=1}^\infty D_n(f)\,,
\ee
where
\be\label{eqn:Dn}
    D_n(f) = \frac{2n}{f}\int^{\zeta_\text{max}}_0d\zeta\int^\infty_0 dz\left\{\frac{\mathcal{S}(z,\zeta)\mathsf n(x(z),\zeta)P_n(\zeta)}{H(z)(1+z)^6t(z)^4}\right\}\,.
\ee
The dilution factor is now explicitly included, and we use $x=L/t$, $L=2n/f'$, $f'=(1+z)f$ to write
\be
    x(z) = \frac{2n}{(1+z)ft(z)}\,.
\ee
We assume a flat $\Lambda$CDM cosmology, with
\be
    H(z) = H_0\sqrt{\Omega_\Lambda + (1+z)^3\Omega_\text{m} + \mathcal{G}(z)(1+z)^4\Omega_\text{r}}\,,
\ee
where we use
\be
    \Omega_\text{m} = 0.308\,,\;\Omega_\text{r} = 9.1476\cdot 10^{-5}\,,\;\Omega_\Lambda = 1-\Omega_\text{m} - \Omega_\text{r}\,,\; H_0 = 100h\,\text{km/s/Mpc}\,,\; h=0.678\,,
\ee
and
\be
    \mathcal{G}(z) = \left(\frac{g_*(z)}{g_*(0)}\right)\left(\frac{g_S(0)}{g_S(z)}\right)^{4/3}\,.
\ee
Here, $g_*$ and $g_S$ are the relative number of relativistic and entropic degrees of freedom, respectively. We use the same numerical implementation of $\mathcal{G}(z)$ as Ref.~\cite{Blanco-Pillado:2017oxo}, courtesy of Masaki Yamada.

\subsection{The GWB from a delta-function loop production function}\label{ssec:dd}

Studies have shown that the loop production function, $\mathsf f(x')$, can be approximated by a delta function $A\delta(x'-x_i)$ with $x_i=0.1$~\cite{Blanco-Pillado:2013qja} and $A$ a normalization constant we will find later. Let's take this approximation and see how it simplifies our calculation of the GWB.

Rewriting Eq.~(\ref{eqn:Dn}) with a delta-function loop production function and the loop number density given by Eq.~(\ref{eqn:n-x-zeta}):
\begin{align}
    D_n(f) = \frac{4An^2\GGmu}{f^2}\int^{\zeta_\text{max}}_0d\zeta\int^\infty_0 dz\left\{\frac{\mathcal{S}(z,\zeta)((1-\chi)\GGmu)^{3-3\nu}P_n(\zeta)}{H(z)(1+z)^7t(z)^5((1-\chi)\GGmu-x(z)\zeta)^{5-3\nu}} \right.\nonumber \\ \left.\delta\left(\frac{\GGmu x(z)}{(1-\chi)\GGmu-x(z)\zeta}-x_i\right)\right\}\,.
\end{align}
Our first step will be to rewrite the Dirac delta function $\delta(x'-x_i)$ in terms of $z$ so that we can perform one of the integrals. With $g(z)=\GGmu x(z)/((1-\chi)\GGmu-x(z)\zeta)-x_i$, $g(\bar z)=0$, we rewrite via $\delta(g(z))\rightarrow \delta(z-\bar z)/|g'(\bar z)|$. This requires us to make use of
\be
    \frac{\partial g}{\partial z} = \frac{(\GGmu)^2(1-\chi)(\partial x/\partial z)}{((1-\chi)\GGmu-x(z)\zeta)^2}\,,\quad \frac{\partial x}{\partial z} = -\frac{2n}{f}\frac{t(z)+(1+z)(\partial t/\partial z)}{(1+z)^2t(z)^2}\,,\quad \frac{\partial t}{\partial z} = -\frac{1}{H(z)(1+z)}\,.
\ee
Additionally, we note that $g(\bar z)=0$ may be rearranged to
\be\label{eqn:z-bar}
    x(\bar z)=\frac{(1-\chi)\GGmu x_i}{x_i\zeta+\GGmu}\,,\qquad (1+\bar z)t(\bar z) = \frac{2n(x_i\zeta+\GGmu)}{(1-\chi)\GGmu fx_i}\,.
\ee
Taking the integral over $z$ and substituting in the above expressions for $x(\bar z)$ and $(1+\bar z)t(\bar z)$ wherever possible yields
\be\label{eqn:Dn-delta}
    D_n(f) = \frac{A(\GGmu)^{3\nu-1}x_i^3f^2}{4n^2}\int^{\zeta_\text{hi}}_0d\zeta\left\{\frac{\mathcal{S}(\bar z,\zeta)(1-\chi)^2P_n(\zeta)}{(1+\bar z)^2(1-t(\bar z)H(\bar z))(x_i\zeta+\GGmu)^{3\nu}}\right\}\,.
\ee
In practice, $\bar z$ is found by numerically solving the second expression in Eq.~(\ref{eqn:z-bar}).

We now have a new upper limit of integration, $\zeta_\text{hi}$, whereas it was previously $\zeta_\text{max}$, the loop normalized age when $\chi=1$. This is because not all $\zeta$ contribute for all $\bar z$. Observe, in the second expression in Eq.~(\ref{eqn:z-bar}), that as $\bar z$ decreases, $\zeta$ must \emph{increase}: fixing $n$ and $f$, the GWs we receive today from very long ago (high $z$) were mostly produced by young loops (low $\zeta$), whereas the GWs we receive today from very recently (low $z$) were mostly produced by old loops (high $\zeta$). However, the lowest $z$ can be is $0$, so there's an upper limit on $\zeta$, defined as the value $\zeta_\text{hi}$ which solves
\be
    t_0 = \frac{2n(x_i\zeta_\text{hi}+\GGmu)}{(1-\chi(\zeta_\text{hi}))\GGmu fx_i}\,.
\ee
Since $\chi(\zeta_\text{max})=1$, $\zeta_\text{hi}<\zeta_\text{max}$ always.

In addition, the monotonic relationship between $z$ and $\zeta$ allows us to split the integral in Eq.~(\ref{eqn:Dn-delta}) in two, based on which cosmological era the loops are emitting in. First, we find $\zeta_\text{eq}$ by solving
\be
    (1+z_\text{eq})t(z_\text{eq}) = \frac{2n(x_i\zeta_\text{eq}+\GGmu)}{(1-\chi(\zeta_\text{eq}))\GGmu fx_i}\,.
\ee
so that the range $0\leq\zeta<\zeta_\text{eq}$ is all loops formed in the radiation era and emitting in the radiation era, and the range $\zeta_\text{eq}<\zeta\leq\zeta_\text{hi}$ is all loops formed in the radiation era and emitting in the matter era. Note that $\zeta_\text{eq}<\zeta_\text{hi}$ always, although for certain parameter choices (particularly large $f/n$) the difference can be negligible.

Accounting for all prefactors, we find
\be\label{eqn:Omega-gw}
    \Omega_\text{gw}(\ln f) = \frac{2A\pi (G\mu)^{1+3\nu} x_i^3f^2}{3H_0^2\Gref^{1-3\nu}}\sum_{n=1}^\infty n^{-2}\int^{\zeta_\text{hi}}_0d\zeta\left\{\frac{\mathcal{S}(\bar z,\zeta)(1-\chi)^2P_n(\zeta)}{(1+\bar z)^2(1-t(\bar z)H(\bar z))(x_i\zeta+\GGmu)^{3\nu}}\right\}\,.
\ee

All that remains is to find the normalization constant. From Eq.~(17) of Ref.~\cite{Blanco-Pillado:2013qja}, we have
\be
    \int^\infty_0\alpha^{3/2}\mathsf f(\alpha)\,d\alpha \approx 1.03\equiv I\,.
\ee
That work defined $\alpha = m/(\mu d_h)$, where $d_h=t/(1-\nu)$ and $m$ is the rest-frame mass of the loop. Thus, setting $m=\mu L$,\footnote{This ignores the energy lost to changing frame; for loops with large $x$, which are the kind we're interested in for the delta-function approximation, the difference is negligible.} $\alpha = (1-\nu)x$. Ref.~\cite{Blanco-Pillado:2013qja} defined $f(\alpha)$ to be the number of loops produced per unit $\alpha$ in volume $d_h^3$ in time $d_h$, so that $\mathsf f(\alpha) = (1-\nu)^{-5}\mathsf f(x)$.  Now changing variables from $\alpha$ to $x$, we find
\be
    \int^\infty_0x^{3/2}\mathsf f(x)\,dx \approx (1-\nu)^{5/2}I\,.
\ee
Thus, with $\mathsf f(x) = A\delta(x-x_i)$, $A = (1-\nu)^{5/2}Ix_i^{-3/2}\approx 5.75$, where we have used the value of $\nu=1/2$ appropriate for the radiation era.

\section{Results}\label{sec:results}

We have applied the methodology described in Sec.~\ref{sec:gwb} to our corpus of backreacted loops described in Sec.~\ref{ssec:corpus} to produce gravitational-wave backgrounds, shown in Fig.~\ref{fig:gwb-detectors} and available for download at \url{https://doi.org/10.5281/zenodo.14036934}~\cite{wachter_2024_14036934}. Tensions run from $\log_{10}(G\mu)=-8.0$ down to $-22.0$, and frequencies span the picohertz to one hundred kilohertz range.

These data represent the most accurate picture to date of the GWB from a local cosmic string network coupled only to gravity. We discuss impacts on detectability, comparisons to prior results, and possible improvements to our methodology below.

\subsection{Comparison to gravitational wave detector sensitivities}\label{ssec:compare-detectors}

One of the great advantages of the cosmic string GWB is that it spans many decades of frequency at amplitudes significant enough to be detected by current or future gravitational-wave telescopes. Related to this broad span, the string GWB carries in it the imprints of historical events in the universe, such as changes to the relativistic degrees of freedom. For this reason, precise measurements of the string GWB are of interest both as evidence for an additional symmetry-breaking stage in the early universe and as a test of standard cosmology.

\begin{figure}
    \centering
    \includegraphics[width=1.00\linewidth]{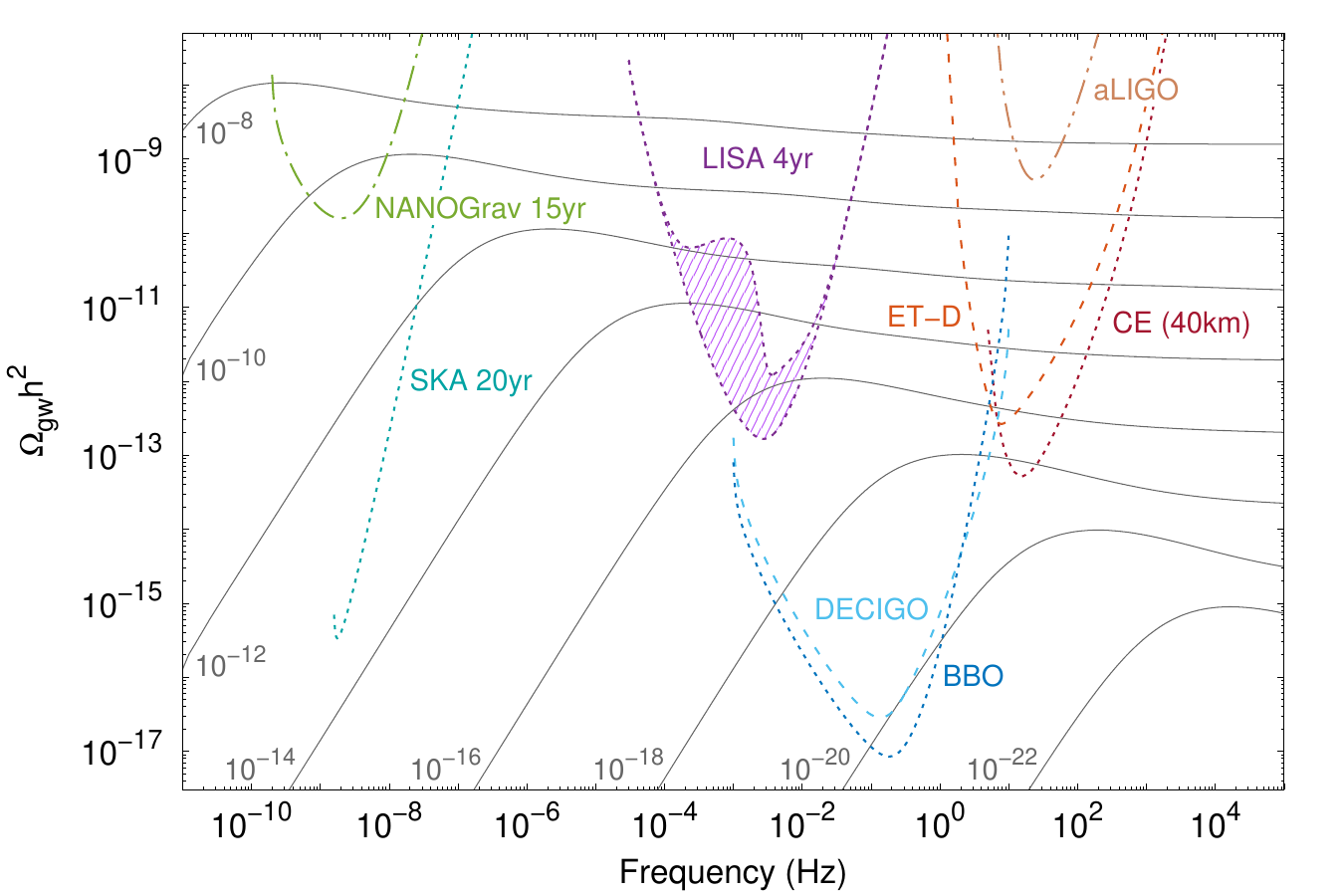}
    \caption{Cosmic string gravitational wave backgrounds (solid grey lines) at various tensions $G\mu$ (indicated by adjacent grey numbers). Several current and future gravitational-wave detector power-law integrated sensitivity curves are overlaid: the NANOGrav 15yr sensitivity (green, dot-dashed); the SKA 20yr sensitivity (cyan, dotted); the LISA 4yr sensitivity (purple, dotted; upper envelope: with astrophysical foregrounds; lower envelope: without foregrounds); DECIGO (light blue, dashed); BBO (dark blue, dotted); the advanced LIGO (aLIGO) design sensitivity (brown, dot-dot-dashed); the Einstein Telescope D (ET-D) configuration (orange, dashed); and the Cosmic Explorer (CE) 40\,km (red, dotted). Many cosmic string GWBs would be visible in multiple detectors simultaneously.}
    \label{fig:gwb-detectors}
\end{figure}

As such, it is informative to consider what cosmic string tensions could be probed through the GWB by current and future GW detectors. Figure~\ref{fig:gwb-detectors}
 summarizes this, comparing GWBs for a wide range of tensions and frequencies against a handful of representative GW detectors' sensitivity curves. String GWBs are shown as grey lines in two-decade steps in tension; current GW detectors' sensitivities are shown in dash-dotted lines, while planned detector sensitivities are shown in dashed or dotted lines. All detector sensitivities are given as power-law integrated sensitivity curves~\cite{Thrane:2013oya} assuming an SNR threshold of 1; unless otherwise specified, the curve is constructed for one year of observing time. We acknowledge~\cite{Schmitz:2020syl} as a major resource for formulae, constants, and references to guide the construction of several of these curves, especially DECIGO and BBO.

The detectors we chose to include are:
\begin{itemize}
    \item For pulsar timing arrays (PTAs):
    \begin{itemize}
        \item NANOGrav, at its 15-year sensitivity~\cite{Agazie_2023,NG15-dataset} to represent the current generation;
        \item the Square Kilometer Array (SKA)~\cite{Carilli:2004nx} at its 20-year sensitivity to represent the next generation.
    \end{itemize}
    \item For space-borne interferometers:
    \begin{itemize}
        \item LISA, at its 4-year sensitivity, where the upper envelope is with astrophysical foregrounds and the lower envelope is without foregrounds~\cite{Babak:2021mhe,Babak:2023lro,Nissanke:2012eh}, to illustrate both foreground noise and represent the (planned) first generation;
        \item DECIGO~\cite{Seto:2001qf} and the Big Bang Observer (BBO)~\cite{Harry:2006fi} to represent the second generation, both assuming a triangular configuration.
    \end{itemize}
    \item For ground-based interferometers:
    \begin{itemize}
        \item Advanced LIGO's design sensitivity (aLIGO)~\cite{LIGO-T1800044-v5} to represent the current generation;
        \item Cosmic Explorer (CE) assuming 40\,km arms~\cite{Srivastava:2022slt} and the Einstein Telescope D configuration (ET-D)~\cite{Hild_2011} to represent the next generation.
    \end{itemize}
\end{itemize}
While precise calculations using the data and tools of each experiment discussed here are necessary for conclusive limits, we can infer from Fig.~\ref{fig:gwb-detectors} some rough bounds on probe-able string tensions. 

Among currently operating experiments, NANOGrav has the greatest reach
for detecting cosmic strings with low $G\mu$.  From
Fig.~\ref{fig:gwb-detectors} it would appear that the NANOGrav 15-year
data would be sensitive to $G\mu$ a few times $10^{-11}$.  If NANOGrav
had seen nothing, indeed strings could have been excluded down to such a
limit.  But NANOGrav did see strong evidence of a GWB signal
\cite{NANOGrav:2023gor}.  Since the source of the signal is not known,
NANOGrav \cite{NANOGrav:2023hvm} could give only an upper limit on
$G\mu$ around $10^{-10}$, assuming both strings and supermassive black
holes are present in the signal.  Similar considerations affect
detecting or setting upper limits on $G\mu$ at other observatories.

As discussed in the next subsection, the effect of the present work is
to reduce somewhat the GWB power for a given $G\mu$, to different
degrees at different frequencies.  This would lead to a small weakening
in observational bounds on cosmic strings, which will be analyzed in
more detail in later work.

A larger PTA like SKA could probe lower string tensions, but this effect
is limited.  At such tensions the GWB frequencies to which PTAs are sensitive
are on the rapidly falling left edge of the cosmic string GWB spectrum.

On 25 January 2024, the LISA mission was officially adopted by ESA with a target launch date of 2035, making LISA the most likely candidate for the next new GW observatory to set bounds on cosmic strings. LISA could measure the cosmic string GWB down to $\approx 10^{-16}$, shown here assuming four years of observing time, consistent with the expected mission duration. The shaded region on the LISA curve indicates where astrophysical foregrounds from galactic and extragalactic compact object mergers are expected to cause confusion noise. If these foregrounds can be subtracted out, LISA might probe string tensions as low as $10^{-17}$~\cite{Auclair:2019wcv,Blanco-Pillado:2024aca}. Confusion noise from multiple GWB sources is an important consideration in general for string detection prospects.

Current ground-based interferometers set a weaker bound at around $G\mu=10^{-9}$. Future ground-based interferometers, such as ET and CE, will be able to set much stronger bounds, competitive with the tensions probed by LISA.

Another nice feature of the broad frequency support of string GWB is the possibility for measuring the signal in multiple detectors at once. E.g., a GWB arising from a network of strings with $G\mu=10^{-14}$ could be seen in LISA, BBO, ET and CE simultaneously. Then, the lower frequencies would probe effects from the radiation-to-matter transition ``bump''  in the GWB, and the higher frequencies probing ``steps'' from changes to relativistic degrees of freedom~\cite{Battye:1997ji,Blanco-Pillado:2017oxo}.\footnote{``Bump'' refers to the spectrum's peak following its low-frequency rising edge, as the peak at around $10^{-8}\,\text{Hz}$ in Fig.~\ref{fig:comparison_1e-10}; ``steps'' refers to the sequence of dips in the spectrum at frequencies higher than the bump, such as begin at around $10^{-2}\,\text{Hz}$ in the same.} Due to the overlap in planned operational lifetimes of the detectors discussed here, as well as others such as TianQin or Taiji, there are many possibilities for studying the string GWB's structure.

\subsection{Comparison to prior results}\label{ssec:compare-prior}

The results presented in this paper provide an update to prior work which modeled backreaction by a Lorentzian convolution process to smooth initially-kinky loops~\cite{Blanco-Pillado:2017oxo}. This convolution process was done to various fractions of the initial loop size, then took the $\Gamma$ of 50\% evaporated loops and assumed that applied for the full lifetime. So, the GWB generated by this method implicitly assumes that all loops are always ``middle-aged''. However, the way backreaction removes power from high modes (small-scale structure) has an important impact on the spectral shape of the GWB, and allowing for larger $\Gamma$ at early times reduces loop lifetimes.

\begin{figure}
    \centering
    \includegraphics[width=0.75\linewidth]{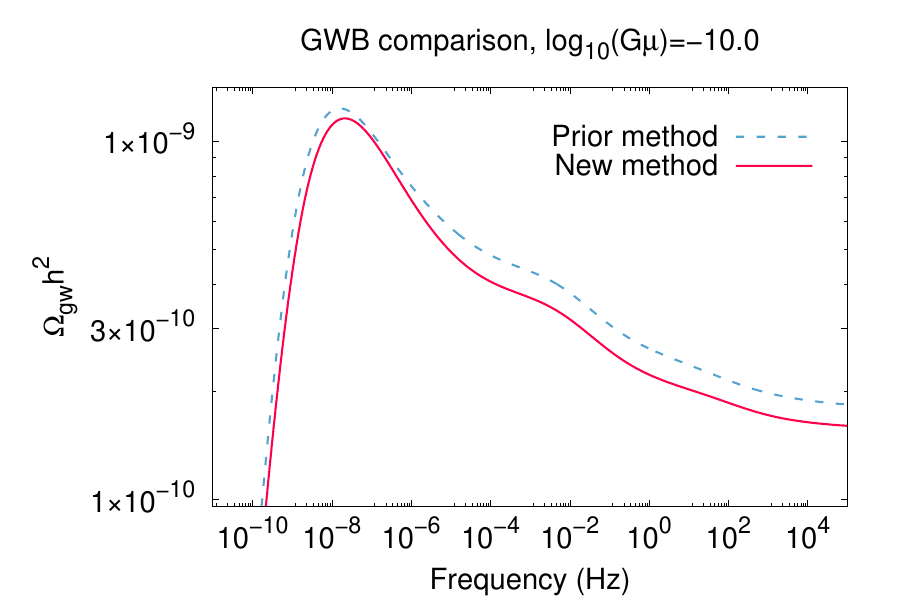}
    \caption{The new method (solid red line) actually calculates how loops evolve under backreaction numerically, while the prior method (dashed blue line) used a toy model of backreaction~\cite{Blanco-Pillado:2017oxo}. The overall effect of using the new, more accurate method is to lower the GWB amplitude across all frequencies.}
    \label{fig:comparison_1e-10}
\end{figure}

Our updated method leads to a moderate reduction in the amplitude of a cosmic string GWB compared to the prior method over the whole range of frequencies; a comparison for a particular tension is shown in Fig.~\ref{fig:comparison_1e-10}.

It is instructive to investigate how this change comes about. There are two principal effects contributing to the difference in the GWBs predicted by the two methods:
\begin{enumerate}
    \item the power emitted in gravitational waves, $\Gamma$, is a constant $\Gref=50$ in the prior method, but in the new method $\Gamma$ is a decreasing function of the loop's age and $\Gamma >\Gref$ for most of the loop's life, and
    \item the typical lifetime of a loop in the new method is less than the typical lifetime of a loop in the prior method.
\end{enumerate}
These two effects are of course not unrelated---the high $\Gamma$ at early ages means that the loops lose length more rapidly, on average, than the prior model with fixed $\Gamma=\Gref$ assumed, even though at later ages the loops have $\Gamma<\Gref$.

A noteworthy effect of the first point above, as elaborated upon in Sec.~\ref{sec:gwb}, is to change the way the power spectrum depends on cosmic time. Older loops look more like the fixed-$\Gamma_*$ loops, but newer loops have much more power at small scales and thus higher frequencies.\footnote{Because $\Gamma\approx\Gamma_*$ at later ages, the Lorentzian convolution captures well the gravitational-wave emission of older loops. Thus the GWB it produces are like the GWB one would expect if only older loops contributed to the GWB.} As a result, the `bump' in the GWB, which is due primarily to loops created in the radiation era but emitting in the matter era, has its peak location shifted to slightly higher frequencies compared to the peak location in the Lorentzian convolution model. This shift, combined with the overall reduction in amplitude due to the shorter loop lifetimes, leads to there being a greater difference between the prior and new methods on the low-frequency side of the GWB's bump compared to the high-frequency side of the bump.

Loops created and emitting in the radiation era span a full range of ages; there are (effectively) always loops which have just been created and loops which are just about to disappear. This is where the second point above becomes most relevant. The radiation-in-radiation contribution to the GWB is a high-frequency plateau, with steps induced by degree-of-freedom changes in cosmic history; because loops live for less time on average than previously assumed, this plateau is decreased in amplitude.

We can estimate the size of this decrease by assuming that we are deep in the radiation era and ignoring degree-of-freedom changes, so we can approximate $H(z)\propto (1+z)^2$, $t(z)\propto (1+z)^{-2}$, $\zeta_\text{hi}\rightarrow\zeta_\text{max}$, and $\mathcal{S}(z,\zeta)=1$. Combining these approximations with Eq.~(\ref{eqn:z-bar}), we arrive at a much-simplified form of Eq.~(\ref{eqn:Omega-gw}):
\be\label{eqn:Omega-simple}
    \Omega_\text{gw} \propto \int^{\zeta_\text{max}}_0 \sqrt{\zeta+\GGmu/x_i}\Gamma(\zeta)d\zeta\,.
\ee
We were able to sum over the $P_n(\zeta)$ to get $\Gamma(\zeta)$ because there was no other $n$-dependence. Since $\GGmu/x_i\ll 1$ for all string tensions we consider, we can neglect it except for very small $\zeta$, and even then we will see it can never be important.

As a basis for comparison, we define $\Omega_\text{gw}^*$ to be $\Omega_\text{gw}$ in the simple case of $\Gamma =\Gref$, giving $(2/3)\Gref$ in  Eq.~(\ref{eqn:Omega-simple}). Thus
\be\label{eqn:Omega-gw-rad}
    \frac{\Omega_\text{gw}}{\Omega_\text{gw}^*} = \frac{3}{2\Gref} \int^{\zeta_\text{max}}_0 \sqrt{\zeta}\Gamma(\zeta)\,d\zeta
\ee
deep in the radiation era.  From Eq.~(\ref{eqn:Gref-integral}), if we had kept $\GGmu/x_i$, its contribution to Eq.~(\ref{eqn:Omega-gw-rad}) could never be more than $(3/2)\sqrt{\GGmu/x_i} \ll 1$.

We can use Eq.~(\ref{eqn:Omega-gw-rad}) to understand the effect of varying $\Gamma$ on the radiation-era plateau. Since we always find that $\Gamma$ is a decreasing function of $\zeta$, we will consider two limiting cases: one where $\Gamma$ is constant, and the other where $\Gamma$ is very high for $\zeta<\zeta_1\ll1$ and then drops immediately to $\Gref$.  In the first case, $\zeta_\text{max} = \Gamma_*/\Gamma$ and $\Omega_\text{gw}/\Omega_\text{gw}^* = \sqrt{\Gref/\Gamma}$.  In the second case, let the additional energy emitted at the beginning be $\Delta\chi = \Gamma \zeta_1$.  The early emission does not contribute to the GWB, because of the $\sqrt{\zeta}$ factor in Eq.~(\ref{eqn:Omega-gw-rad}).  But it reduces $\zeta_\text{max}$ to $1-\Delta\chi$, giving $\Omega_\text{gw}/\Omega_\text{gw}^* = (1-\Delta\chi)^{3/2}$.  Thus in one limit we have the power 1/2 and any other 3/2, so we might expect that in real cases the effect will be intermediate.

\begin{figure}
    \centering
    \includegraphics[width=0.75\linewidth]{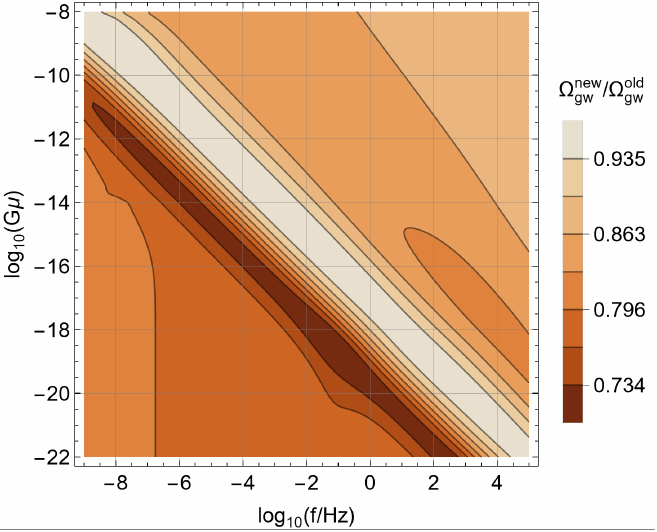}
    \caption{The new method for calculating the string GWB produces a lower amplitude $\Omega_\text{gw}$ at all tensions and frequencies considered in comparison to the prior method. This difference, expressed as the ratio of the new to old amplitude, ranges from $0.97$ down to $0.71$, with a median of $0.83$. The difference is most pronounced just below the bump in the GWB, and least pronounced just above that bump (cf. the diagonal whitish band above, and the horizontal shift of the bump in Fig.~\ref{fig:gwb-detectors} as $G\mu$ changes). At frequencies, high above the bump, the difference approaches a factor $\zeta_\text{max}$.}
    \label{fig:comparison_full-range}
\end{figure}

To compute it, we use $d\chi/d\zeta = \Gamma(\zeta)/\Gref$ to transform the integral, giving
\be\label{eqn:sqrtzeta}
    \frac{\Omega_\text{gw}}{\Omega_\text{gw}^*}= \frac32\int^1_0 \sqrt{\zeta(\chi)}\,d\chi\,.
\ee
Using the $\chi(\zeta)$ function we found from our numerical study to obtain $\zeta(\chi)$, we integrate and find that the GWB amplitude at the radiation-era plateau is reduced by a factor $\approx 0.87$. This is close to $\zeta_\text{max}\approx 0.88$, but the detailed agreement is likely coincidental. We see in Fig.~\ref{fig:comparison_full-range} that the difference in methods is approaching the $\approx 0.87$ limit at high frequencies. 

The largest difference in methods in terms of percent change in fact happens at frequencies just below the ``bump'' in the GWB. We see the least difference at frequencies at and just above the bump, where the amplitudes nearly match before the new method's $\Omega_\text{gw}$ drop more rapidly.

Because an effect of the shorter lifetimes of loops compared to previous predictions is a decreased amplitude of the radiation-era plateau, there are regions in $G\mu$--$f$ space where the ratio of the new to old GWBs is effectively constant. Or, the prediction made using the new method is (nearly) a downwards translation of the prediction made in the old model (cf. the prior and new curves above $f\sim 10^{-4}\,\text{Hz}$ in Fig.~\ref{fig:comparison_1e-10}). Decreasing $G\mu$ also leads to a decrease in amplitude, and so we can inquire if a detection pipeline using the GWB made with the old method and one using the GWB made with the new method could report similar levels of confidence in a detection, with the first pipeline reporting a lower tension than the second. If the range of frequencies probed is very small, or the discrepancy in reported tensions very small, this remains possible; however, note that decreasing $G\mu$ also shifts the general features of the GWB to higher frequencies (and induces small changes in those features). Thus with a sufficient range of frequency information, and at high enough sensitivity, the GWB predicted by the new method should not be mistaken for the GWB predicted by the prior method.

While our focus has been so far on comparing the new method to the prior method, other methods are common in the study of cosmic string GWBs. We comment briefly on two of the most popular, the pure cusp and pure kink power spectrum models, in Appendix~\ref{app:cusp-kink}.

\subsection{Discussion}\label{ssec:discuss}

These results represent the most accurate calculation of cosmic string GWB to date. That said, there is always room for improvement. Let us discuss possible adjustments and extensions to the approach laid out above.

One of the fundamental effects we have reported on in this work and its companion piece~\cite{results-paper} is that loops, which are produced with large amounts of small-scale kinky structure (``wiggliness''), quickly shed this wiggliness via gravitational backreaction. Our results track how the spectrum and gravitational power emission evolve with backreaction for loops taken from a network produced in a simulation of a symmetry-breaking transition which then evolves in a standard universe, producing non-self-intersecting loops. These loops which we take from these simulations have hundreds to thousands of kinks. Some kinks are produced by reconnections during the simulation, but others remain from the initial conditions.  Real loops do not have kinks remaining from the time of string formation, but they have even more kinks than loops from simulations, due to a longer dynamic time range over which the network has evolved.  A simple estimate is that structure on long strings is smoothed at scales below $\Gamma_l G\mu t$, so this is the typical distance between kinks.  Here $\Gamma_l$ is the radiation rate on long strings.  With current observational constraints, $\Gamma_l G\mu<10^{-8}$.  A typical string is formed with length $0.1t$, so we would expect it to have about $10^7$ kinks.  To obtain such loops would require computational resources beyond any current computer. Similarly, evolving a representative population of these loops under backreaction is computationally a no-go. As such, we must estimate how a real loop would change at very low $\zeta$ values, before it's been smoothed to a level such that our loop results apply.

We model the impact of early emission on the GWB as follows. Assume that all of the extra length is lost immediately after a loop's creation, on timescales far below $\zeta_\text{max}$. This is essentially the second model considered above, except that we're starting from a $\Gamma$ that is not constant.  The length lost introduces a $\Delta\chi$ additive correction to our $\chi(\zeta)$ function.  Deep in the radiation era with no changes to degrees of freedom, the effect is the same as above, leading to a change in $\Omega_\text{gw}$ by a factor $(1-\Delta\chi)^{3/2}\approx 1-(3/2)\Delta\chi$ for small $\Delta\chi$.

The real situation is more complicated and must be handled using the $\chi$-$\zeta$ relationship we found from our numerical study. We added various early bursts of radiation---various values of $\Delta\chi\ll 1$---into the calculation of Eq.~(\ref{eqn:Omega-gw-rad}) and found the result was consistently lowered by a factor approximately $1-(3/2)\Delta\chi$, except that the $3/2$ was modified by less than 10\%.

So, although naturally occurring loops may have different structure at the smallest scales from those modeled here, in both cases such structures are eliminated during the early stages of evolution through gravitational wave emission. The total energy emitted through this process is small, leading to a limited impact on the resulting GWB.

A second correction to consider is the impact of more significant self-interactions on the GWB. The sub-population we used to construct the GWB reported on here experienced ``no major self-intersections'', by which we mean that less than 5\% of the original loop's length is lost to looplet production over the course of its life we studied. A smaller number of loops experienced ``major self-interactions'' (the conjugate population evolved to 70\% total length lost). Apart from these self-interactions, which appear on a $\chi$ vs. $\zeta$ plot as discontinuous upwards jumps, these loops evolved in the same way as their no-majors counterparts.

The net effect of self-intersections is then to reduce the maximum age of a loop compared to a loop without self-intersections. Accounting for this should serve to further lower the GWB. To make a first estimate, we observe that the fraction of length lost to self-intersections is roughly log-normally distributed (see Fig.~4(c) of~\cite{results-paper}). Comparing the log-median of this length loss for the no-majors subpopulation to the log-median of the length loss for the full 70\%-evaporated population, we find $\approx 0.010$ vs. $\approx 0.024$. Thus, self-intersections could lower $\zeta_\text{max}$ by approximately another $0.014$, and the GWB would be further reduced by some similarly small fraction, following the analysis of early emission above.

Finally, we note that we have not included the contributions to the GWB of loops produced and emitting in the matter era. This is due to our backreacted loops being produced in radiation-era simulations; applying results for how they evolve to loops produced in the matter era might not be correct. For the range of tensions we study, the ``matter-in-matter'' contribution should be sub-dominant at the frequencies of interest shown in this work (see Fig.~14 of Ref.~\cite{Blanco-Pillado:2013qja}). However, the GWB's power-law decay for frequencies very far below the ``bump'' may be steeper than they would be if the matter-in-matter contribution were included.

\section{Conclusions}\label{sec:conc}

Accounting for backreaction on cosmic string loops by numerical evolution, rather than by toy models, changes the standard picture of cosmic string gravitational wave backgrounds. Because the loop power emission coefficient $\Gamma$ is larger for younger loops, backreacted loops live shorter lives in comparison to loops evolving with a fixed, ``average'' $\Gamma$, which causes the GWB to be lower at all frequencies and for any tension.

This reduction in $\Omega_\text{gw}$ varies in frequency and tension, so applying a uniform reduction in amplitudes is not a sufficiently complex model for the effects of backreaction. It is likewise not correct to simply update constant-$\Gamma$ models to use the average of age-varying $\Gamma$; for one, this incorrectly predicts the reduction in the radiation-era plateau to go like $\zeta_\text{max}^{1/2}$ rather than the observed $\approx \zeta_\text{max}$, and for another, the effect of changing $\Gamma$ in a constant-$\Gamma$ model is to scale and translate the GWB mostly rigidly, inconsistent with the changes seen here.

Because the change to $\Omega_\text{gw}$ is a reduction of at most 30\%, we do not anticipate existing bounds on cosmic strings set by gravitational-wave non-observations to change in an important way. However, future studies of string GWB detectability should use the updated curves, with backreaction included, for greater precision.  Perhaps most importantly, we now know the correct GWB spectrum from cosmic strings (to within a few percent, as discussed in Sec.~\ref{ssec:discuss}).  So it is not necessary to use several different possible spectra as was done in Refs.~\cite{Blanco-Pillado:2021ygr,NANOGrav:2023hvm}.

To encourage the usage of this model, we have made tabular data for the new cosmic string GWB discussed here available at \url{https://doi.org/10.5281/zenodo.14036934}~\cite{wachter_2024_14036934}. The frequency range is from $f=10^{-12}\,\text{Hz}$ to $f=10^5\,\text{Hz}$ with a logarithmic spacing of $0.01$, and the tension range is from $G\mu=10^{-8}$ down to $G\mu=10^{-22}$ with a logarithmic spacing of $0.1$.

\section{Acknowledgments}

K. D. O. was supported in part by National Science Foundation Grant Nos. 2111738 and 2412818. J. J. B.-P. is supported by the PID2021-123703NB-C21 grant funded by the Ministerio de Ciencia, Innovaci\'on y Universidades and Agencia Estatal de Investigaci\'on, MCIN/ AEI /10.13039/501100011033/; by the European Regional Development Fund, ``A way of making Europe''; by the Basque Government grant (IT-1628-22); and by the Basque Foundation for Science (IKERBASQUE). The authors acknowledge the Tufts University High Performance Computing Cluster~\cite{hpc}, which was utilized for the research reported in this paper.

\appendix

\section{Recovering the previous loop number density}\label{app:original-n}

If we neglect the effects of changing $\Gamma$, we should recover the results of Ref.~\cite{Blanco-Pillado:2013qja} for the loop number density:\footnote{See Eq.~(15) of Ref.~\cite{Blanco-Pillado:2013qja}; they use $\alpha=(1-\nu)x$ as the scaling loop size measure.}
\be
    \mathsf n(x) = \frac{\int^\infty_x(x'+\GGmux)^{3-3\nu}\mathsf f(x')\,dx'}{(x+\GGmux)^{4-3\nu}}\,.
\ee
To show this consistency, we will integrate Eq.~(\ref{eqn:n-x-zeta}) with respect to $\zeta$ in the case of constant $\Gamma=\Gref$, which implies $\chi=\zeta$, and $\zeta_\text{max}=1$. In addition, we use
\be
    x' = \frac{\GGmux x}{(1-\zeta)\GGmux-x\zeta}\,,
\ee
the argument of the loop production function $\mathsf f$ in Eq.~(\ref{eqn:n-x-zeta}), to rewrite the age as
\be
    \zeta = \frac{\left(1-x/x'\right)\GGmux}{x+\GGmux}\,.
\ee
We can then change variables, noting that $\zeta=0$ is achieved when $x=x'$ and $\zeta=1$ is achieved when $x=0$, for which $x'$ can be any positive value---i.e., we must integrate up to infinite $x'$ to capture all $\zeta=1$. So,
\be\label{eqn:n-x-integral}
    \mathsf n(x) = \int^1_0\mathsf n(x,\zeta)\,d\zeta = \int^\infty_x\mathsf n(x,x')\left(\frac{d\zeta}{dx'}\right)\,dx' = \int^\infty_x\mathsf n(x,x')\left(\frac{\GGmux x}{x'^2(x+\GGmux)}\right)\,dx'\,.
\ee
We can also write
\be
    1-\zeta = \frac{x}{x'}\left(\frac{x'+\GGmux}{x+\GGmux}\right)\,,
\ee
and then
\be
    \mathsf n(x,x') = \frac{1}{\GGmux}\frac{x'^2}{x}\left(\frac{x'+\GGmux}{x+\GGmux}\right)^{3-3\nu}\mathsf f(x')\,.
\ee
Inserting this into Eq.~(\ref{eqn:n-x-integral}), we get
\begin{align}
    \mathsf n(x) &= \int^\infty_x \left[\frac{1}{\GGmux}\frac{x'^2}{x}\left(\frac{x'+\GGmux}{x+\GGmux}\right)^{3-3\nu}\right]\left[\frac{\GGmux x}{x'^2(x+\GGmux)}\right]\mathsf f(x')\,dx'\nonumber\\ &= \frac{\int^\infty_x(x'+\GGmux)^{3-3\nu}\mathsf f(x')\,dx'}{(x+\GGmux)^{4-3\nu}}\,,
\end{align}
precisely as desired.

\section{Relic loop densities}\label{app:relic-loops}

A significant source of the structure in the GWB at low frequencies are loops formed in the radiation era but emitting in the matter era. For $t>t_\text{eq}$, the number density of these loops is changed only by the differing rate of expansion.  In addition, a proper calculation of the loop density for loops created and emitting in the radiation era requires us to propagate the loop density at the time of creation to the time of emission, taking into account changes in the number of degrees of freedom.

In both of these cases, once loops of a particular size are no longer being formed in significant quantities, their number density changes only via dilution due to cosmic expansion and shrinkage due to gravitational-wave emission.  Consider a population of loops with some density $\mathsf n(x_1,\zeta)$ at an arbitrary time $t_1$. There are a range of loops of sizes $dx_1$ and ages $d\zeta$ in a volume $t_1^3$, and thus per comoving volume like $(a_1/t_1)^3\mathsf n(x_1,\zeta_1)dx_1 d\zeta_1$. Assuming loop production at sizes $x_1$ ceases at time $t_1$, the number density of loops per comoving volume does not change, and so at some time $t>t_1$ we can say
\be
    \left(\frac{a}{t}\right)^3\mathsf n(x,\zeta)dxd\zeta = \left(\frac{a_1}{t_1}\right)^3\mathsf n(x_1,\zeta_1)dx_1 d\zeta_1\,.
\ee
Since we want this analysis to be agnostic as to the era of $t$, we use $a\propto 1/(1+z)$ to rewrite the scale factors.\footnote{In particular, for finding the contribution of relic radiation loops in the matter era, making the substitution $a\propto t^\nu$ underestimates the size of the ``bump'' in the GWB.} So,
\be
    \mathsf n(x,\zeta;t>t_1) = \left(\frac{t}{t_1}\right)^3\left(\frac{1+z}{1+z_1}\right)^3\mathsf n(x_1,\zeta_1)\left|\mathcal{J}(x,\zeta)\right|\,,
\ee
where $\mathcal{J}(x,\zeta)$ is the Jacobian transformation going from $(x,\zeta)$ to $(x_1,\zeta_1)$. To find this, we first use Eqs.~(\ref{eqn:L0-t0}) to write
\be
    x_1 = x\left(\frac{t}{t_1}\right)\left(\frac{1-\chi_1}{1-\chi}\right)\,,\qquad \zeta_1 = \zeta + \left(\frac{(1-\chi)\GGmu}{x}\right)\left(\frac{t_1}{t}-1\right)\,.
\ee
Then, using $\chi_1=\chi(\zeta_1(\zeta,x))$, we compute the Jacobian and find
\be
    \mathsf n(x,\zeta;t>t_1) = \left(\frac{t}{t_1}\right)^4\left(\frac{1+z}{1+z_1}\right)^3\left(\frac{1-\chi_1}{1-\chi}\right)\mathsf n(x_1,\zeta_1)\,.
\ee
Rewriting $\mathsf n(x_1,\zeta_1)$ to be in terms of $x$ and $\zeta$ proceeds as follows. First, observe that
\be
    x_1\zeta_1 = (1-\chi_1)\GGmu\left(1-\frac{1}{1-\chi}\frac{x}{x'}\frac{t}{t_1}\right)\,.
\ee
Then, from Eq.~(\ref{eqn:n-x-zeta}) and $x'/x = (L't)/(Lt')$, and setting $\nu=1/2$ (as we are always producing loops in the radiation era), we obtain
\be
    \mathsf n(x_1,\zeta_1) = \left(\frac{t_1}{t}\right)^{5/2}\left(\frac{1-\chi}{1-\chi_1}\right)\left[\frac{x}{(1-\chi)^2\GGmu}\left(\frac{t}{t'}\right)^{7/2}\mathsf f(x')\right]\,.
\ee
Recalling that $t/t'=(1-\chi)\GGmu/((1-\chi)\GGmu-x\zeta)$ and $x'=L'/t' = \GGmu x/((1-\chi)\GGmu-x\zeta)$, we see that the term in square brackets is just Eq.~(\ref{eqn:n-x-zeta}) with $\nu=1/2$. So,
\be
    \mathsf n(x_1,\zeta_1) = \left(\frac{t_1}{t}\right)^{5/2}\left(\frac{1-\chi}{1-\chi_1}\right)\mathsf n(x,\zeta)
\ee
and \emph{in toto}
\be
    \mathsf n(x,\zeta;t>t_1) = \left(\frac{t}{t_1}\right)^{3/2}\left(\frac{1+z}{1+z_1}\right)^3\mathsf n(x,\zeta) = \mathcal{S}(z,z_1)\mathsf n(x,\zeta)\,,
\ee
where $\mathcal{S}$ is the dilution factor between $z_1$ and $z$. This general result is applied in two slightly different ways, depending on the era in which the loops are emitting.

For radiation-era loops emitting in the radiation era, $t_1$ and $z_1$ are the time and redshift of the formation of the loops, which we'll denote by $t'$ and $z'$. To determine the time of formation, recalling the definition of $\zeta$, we see that $t'=\GGmu t/(x'\zeta+\GGmu)$. Since $x'=x_i$ is a fixed value in our delta-function approximation of the loop production function, and since we have functions relating $t$ and $z$, given any $(z,\zeta)$ pair, we can recover $t'$ and $z'$.

For radiation-era loops emitting in the matter era, we take $t_1$ and $z_1$ to be the time and redshift of loop formation, as before. Now, we should additionally constrain ourselves to only look at loops which formed, at the latest, at $t_\text{eq}$. This means that for any given $(z,\zeta)$ pair, with $t(z)$ the time of emission, we write
\be
    \frac{\GGmu t(z)}{\zeta x_i + \GGmu} \leq t_\text{eq}
\ee
as the condition for a radiation-era loop to contribute in the matter era.

Since redshift at formation can always be found knowing the redshift at emission and the age of the loop, when we take the starred values to be the initial values, we will write the scaling factor as a function of redshift and age, i.e., $\mathcal{S}(z,\zeta)$.

As a final note, the time-of-formation procedure can be used to show that loops formed at the end of the radiation era which still exist today must have a normalized age of at least $\zeta_\text{min}=(\GGmu/x_i)(t_0/t_\text{eq}-1)$. Assuming $x_i=0.1$ and setting $\zeta_\text{min}=\zeta_\text{max}$, we see that strings with $G\mu \gtrsim 6.5\cdot 10^{-9}$ cannot have radiation-era loops survive to the present day.

\section{The distribution of loops by normalized age}\label{app:zeta-distribution}

Since it is the youngest loops which have power spectra most different from the canonical picture of loops, we may be interested to learn something about their relative abundance. 

Assume that all loops are produced at the same scaling fractional size, $\mathsf f(x') = A\delta(x'-x_i)$. Then, we can integrate $\mathsf n(\zeta,x)$ over all allowed $x$ to find $\mathsf n(\zeta)$. This does require us to rewrite the Dirac delta function:
\begin{subequations}\begin{align}
    \delta\left(\frac{\GGmu x}{(1-\chi)\GGmu-x\zeta}-x_i\right) &= \frac{(1-\chi)(\GGmu)^2}{\left(x_i\zeta+\GGmu\right)^2}\delta(x-\bar x)\,,\\\bar x &= \frac{(1-\chi)\GGmu x_i}{x_i\zeta+\GGmu}\,.
\end{align}\end{subequations}
Integrating:
\be
    \mathsf n(\zeta) = \int^\infty_0 \mathsf n(\zeta,x)\,dx = A(\GGmu)^{1+3\nu}x_i\left(x_i\zeta+\GGmu\right)^{2-3\nu}\,.
\ee
This has a power-law dependence on $\zeta$ for $\zeta\gg x_i/\GGmu$, but is cut off at low $\zeta$ (very young loops), asymptoting towards a value of $Ax_i/\GGmu$ rather than going to zero. Or, there is always some non-negligible number of loops ``fresh from the network''.

Finally, we can use our expression for $\mathsf n(\zeta)$ to define a probability density function for $\zeta$ by the usual method:
\be
    \operatorname{Pr}(\zeta) = \frac{\mathsf n(\zeta)}{\int^{\zeta_\text{max}}_0\mathsf n(\zeta')\,d\zeta'} = \frac{(3-3\nu)x_i\left(x_i\zeta+\GGmu\right)^{2-3\nu}}{\left(x_i\zeta_\text{max}+\GGmu\right)^{3-3\nu}-(\GGmu)^{3-3\nu}}\,.
\ee
A brief observation about this probability distribution: for $x_i\zeta\gtrsim \GGmu$, $\operatorname{Pr}(\zeta)\propto \zeta^{2-3\nu}$, but for $x_i\zeta\lesssim \GGmu$, $\operatorname{Pr}(\zeta)\propto (\GGmu/x_i)^{2-3\nu}$. This is consistent with the earlier observation that there are always some loops which have just been produced. However, since $x_i\ggg \GGmu$ for cosmologically-relevant loops, very young loops are suppressed compared to middle-aged or old loops.

\section{Further comparison to commonly-used power spectra}\label{app:cusp-kink}

\begin{figure}
    \centering
    \includegraphics[width=0.75\linewidth]{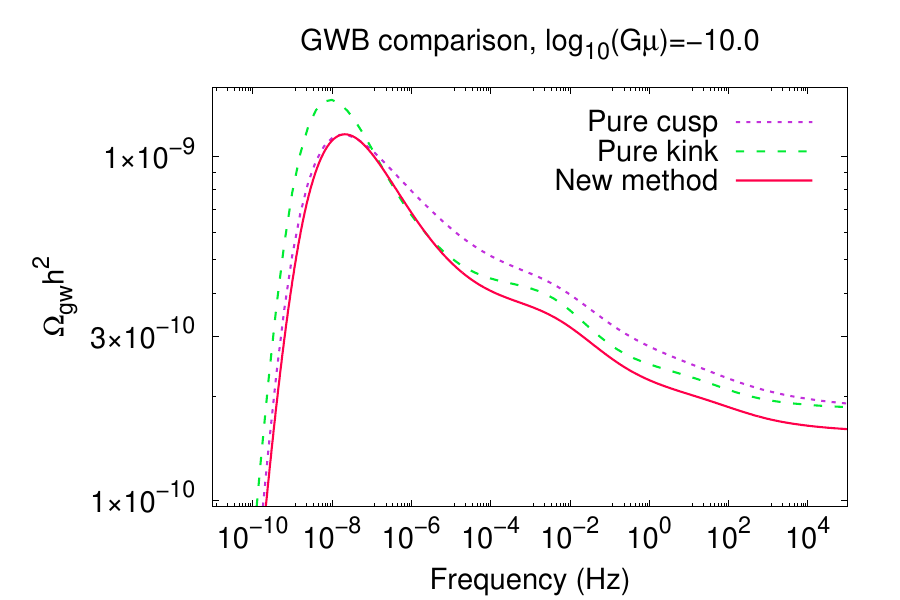}
    \caption{Gravitational-wave backgrounds at via the new method (solid red line), and using pure cusp (dotted purple line) and pure kink (dash-dotted green line) power spectra. These methods overestimate the GWB at high frequencies, but the difference in power spectrum shapes lead to potential overlaps at mid to low frequencies.}
    \label{fig:cusp-kink}
\end{figure}

A major goal of this work has been to improve upon prior modeling of the cosmic string GWB, as laid out in Ref.~\cite{Blanco-Pillado:2017oxo}, and as such we have discussed our new results in comparisons to the ones found there. However, it is popular in the cosmic string community to use the so-called ``pure cusp'' and ``pure kink'' power spectra (see, e.g., ~\cite{Auclair:2019wcv}), which have the form
\begin{equation}
    P_n = \frac{\Gamma}{\zeta(q)}n^{-q}\,,
\end{equation}
with $q=4/3$ for pure cusp and $q=5/3$ for pure kink. For completeness, we present a comparison of GWBs generated using pure cusp and pure kink power spectra and fixed, constant $\Gamma=50$ to the GWB generated using our new method, shown in Fig.~\ref{fig:cusp-kink}.
At high frequencies, the new method GWB is smaller than the prior method GWBs by a factor $\approx 0.87$, for the reasons discussed in Sec.~\ref{ssec:compare-prior}. At lower frequencies, some interesting overlaps emerge; around the microhertz range, the pure kink GWB closely matches the new method GWB, and at and below the bump, the pure cusp GWB closely matches the new method GWB. As far as we are aware, these are coincidences. While we have illustrated this for a single tension of $G\mu=10^{-10}$, the same general overlaps should recur at any tension due to the shifting and scaling of the GWB. Attempts at detection should be able to distinguish these GWB given sufficient frequency range and sensitivity, but certain lower frequencies may be more challenging than higher frequencies.

\bibliography{paper}

\end{document}